\documentclass[aps,prl,amsmath,amssymb,superscriptaddress,footinbib,longbibliography,twocolumn]{revtex4-2}
\usepackage{graphicx}
\usepackage[colorlinks,linkcolor=blue,anchorcolor=blue,citecolor=blue,urlcolor=black]{hyperref}
\usepackage{amsmath}

\usepackage{color}
\usepackage{soul}
\usepackage{ulem}

\usepackage{lineno}

\begin{document}

\title{Observation of quantum multi-Mpemba effect in a trapped-ion system}

\author{Gang Xia}
\thanks{These authors contributed equally to this work.}
\affiliation{College of Science, National University of Defense Technology, Changsha 410073, China}

\author{Yu-Jie Zheng}
\thanks{These authors contributed equally to this work.}
\affiliation{College of Science, National University of Defense Technology, Changsha 410073, China}

\author{Jing Huang}
\affiliation{College of Science, National University of Defense Technology, Changsha 410073, China}

\author{Chun-Wang Wu}
\affiliation{College of Science, National University of Defense Technology, Changsha 410073, China}

\author{Yi Xie}
\affiliation{College of Science, National University of Defense Technology, Changsha 410073, China}
\affiliation{Hefei National Laboratory, Hefei 230088, Anhui, China}

\author{Ting Chen}
\affiliation{College of Science, National University of Defense Technology, Changsha 410073, China}
\author{Wei Wu}
\affiliation{College of Science, National University of Defense Technology, Changsha 410073, China}
\affiliation{Hefei National Laboratory, Hefei 230088, Anhui, China}

\author{Weibin Li}
\affiliation{School of Physics and Astronomy, and Centre for the Mathematics and Theoretical Physics of Quantum Non-equilibrium Systems, University of Nottingham, Nottingham NG7 2RD, United Kingdom}

\author{Hui Jing}
\affiliation{College of Science, National University of Defense Technology, Changsha 410073, China}
\affiliation{Key Laboratory of Low-Dimensional Quantum Structures and Quantum Control of Ministry of Education, Department of Physics and Synergetic Innovation Center for Quantum Effects and Applications, Hunan Normal University, Changsha 410081, China}

\author{Jie Zhang}\email{zj1589233@126.com}
\affiliation{College of Science, National University of Defense Technology, Changsha 410073, China}

\author{Yan-Li Zhou}\email{ylzhou@nudt.edu.cn}
\affiliation{College of Science, National University of Defense Technology, Changsha 410073, China}

\author{Ping-Xing Chen}\email{pxchen@nudt.edu.cn}
\affiliation{College of Science, National University of Defense Technology, Changsha 410073, China}
\affiliation{Hefei National Laboratory, Hefei 230088, Anhui, China}
\date{\today}

\begin{abstract}
The quantum Mpemba effect (ME) in Markovian systems is conventionally explained by a smaller overlap between the initial state and the slowest decay mode (SDM). Such state, initially farther away from equilibrium or steady state, relaxes faster than closer ones, resulting to a crossing of their trajectories. This picture, by neglecting the transient dynamics, holds in the long-time limit. Here we experimentally observe multiple trajectory crossings (multi-ME) in the relaxation dynamics of a trapped ion. Such novel dynamics takes place in a unusual scenario where the initial state instead has a larger overlap with the SDM. We develop a theoretical framework based on relaxation speed to understand the multi-ME. We show that the initial relaxation speed is governed by the fastest decay mode, which together with the SDM overlap gives a phase diagram that reveals both the occurrence and the types of quantum ME observed in our experiment. Our study goes beyond the simple picture based on the long-time limit, tracks continuously the quantum ME dynamics, and establishes a comprehensive framework to describe the transient quantum relaxation. 
\end{abstract}
 
\maketitle

\begin{figure*}[htb]
	\includegraphics[scale=0.9]{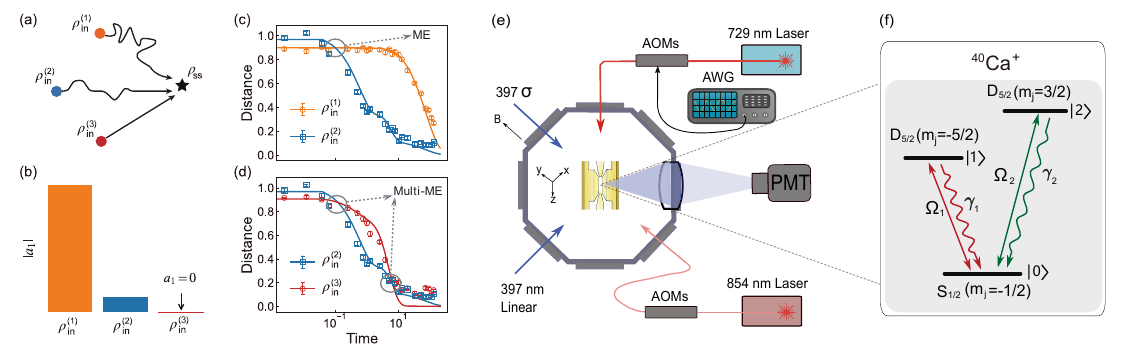}
	\caption{(a) Initial state dependent relaxation process. (b) The overlaps $|a_1|$ between the initial states $\rho_{in}^{1,2,3}$ and the SDM. $a_1=0$ corresponds to the strong ME. (c) The quantum ME is witnessed by a single crossing of distance curves: the initial state $\rho_{in}^{(2)}$ starts a greater distance from $\rho_{ss}$ than the initially closer state $\rho_{in}^{(1)}$ but stabilizes first, for the reason having a smaller overlap $|a_1|$. (d) The multi-ME is witnessed by two crossings and can not be understated by smaller overlap $|a_1|$ anymore. (e) Trapped ion setting. One Zeeman sublevel in the ground state and two Zeeman sublevels in D state are coupled by a bichromatic 729 nm laser beam with linewidth about 10 Hz. The two transitions are simultaneously stimulated by driving an acoustic-optic modulator (AOM) via an arbitrary waveform generator (AWG) with two RF frequencies. (f) The relevant energy levels of a single $^{40}$Ca$^{+}$ ion involved in the experiment. The parameters are $\Omega_1 = 2\pi \times 20.0 \ \text{kHz}, \Omega_2 = 0.06\,\Omega_1, \gamma_1 = 2\pi \times 40.0 \ \mathrm{kHz}, \gamma_2 = 2\pi \times 0.03 \ \mathrm{kHz}$.}
	\label{fig:scheme}
\end{figure*}

\textit{Introduction}—Predicting and controlling relaxation processes in nonequilibrium systems remains a fundamental challenge in modern physics \cite{Teza2026,Ares2025,Pemartin2024}, with profound implications for quantum technologies and complex systems \cite{Seifert2012,Gong2022,Beato2025,Bao2025,Sun2024}. Central to this challenge is the exploration of anomalous relaxation phenomena, notably the Mpemba effect (ME) \cite{Mpemba1969, Bechhoefer2021,Lasanta2017,Lu2017,Klich2019,Kumar2017,Yang2020,Zhang2022PRE,Biswas2023,Teza2023,Nava2025,Van2025}, where a system initialized farther from equilibrium or stationary can relax faster than one starting closer (see Fig. \ref{fig:scheme} (a,c)). Within the framework of Markovian open quantum systems, the relaxation structures can exhibit metastability and a hierarchy of relaxation timescales \cite{Macieszczak2016a,Rose2016,Ivander2023,Minganti2018,Macieszczak2017}, thereby giving rise to much richer relaxation behaviors, such as the quantum ME \cite{Carollo2021, Kochsiek2022, Ares2025,Chatterjee2023, Nava2024,Ivander2023} and strong ME \cite{Longhi2026, Zhang2025}. 
A pivotal theoretical explanation posits that this and related phenomena arises when the farther initial state has a smaller overlap with the slowest decay mode (SDM) of the relaxation dynamics \cite{Lu2017,Klich2019,Carollo2021,Bechhoefer2021} (see Fig. \ref{fig:scheme}(b)).

This criterion successfully frames the quantum ME as a finite-time crossing of relaxation trajectories from different initial conditions \cite{Bechhoefer2021,Kumar2022,Ares2023,Rylands2024} and has been validated across both classical \cite{Kumar2020b,Kumar2022,Ibanez2024,Tian2025} and quantum experimental platforms \cite{Shapira2024,Zhang2025,Joshi2024,schnepper2025}. Indeed, the prevailing explanation for the ME in Markovian systems, which centers on the role of the SDM, naturally emphasizes long-time limit behavior and offers a clear criterion for determining the final relaxation order. However the remaining question is, can quantum ME emerge if the farther initial state has large overlaps with the SDM? if so, what is the origin that drives the ME? 

In this work, we experimentally observe a such quantum ME characterized by two crossings of the distance curves, which we term multi‑ME ~\cite{Chalas2024, Zhang2026}, occurring in the relaxation dynamics of a trapped ion. As shown in Fig.~\ref{fig:scheme}(a,b), for a pair of states \(\rho_{\mathrm{in}}^{(2)}\) and \(\rho_{\mathrm{in}}^{(3)}\), the farther initial state \(\rho_{\mathrm{in}}^{(2)}\) has a larger overlap with the SDM, i.e., \(|a_1^{(2)}| > |a_1^{(3)}|\). Surprisingly, even for the strong ME initial state where \(|a_1^{(3)}| = 0\), we still witness two crossings of the distance curves: the initial state \(\rho_{\mathrm{in}}^{(2)}\) starts at a greater distance from \(\rho_{\mathrm{ss}}\) and relaxes faster in the beginning, but is then overtaken by \(\rho_{\mathrm{in}}^{(3)}\) at a later time (see Fig.~\ref{fig:scheme}(d)). We use the relaxation speed as a predictive and intuitive tool to explain the transient dynamics in the quantum multi-ME that goes beyond the long-time limit. It is able to capture the combined influence of the initial state and the generator of the master equation \cite{Nava2025,Deffner2017,Deffner2017a}. We show that the initial relaxation speed is governed by the fastest decay mode, which together withe the SDM overlap, reveals both the occurance and the type of the quantum ME. This speed monitors the instantaneous relaxation process, while it requires no prior knowledge of the trajectory distance.  As a result, this work establishes a dynamical framework for transient relaxation in open quantum systems, with implications for optimal state preparation and many-body dynamics simulations.

\textit{Model}—We investigate dynamics of a open quantum sytem with a Lindblad master equation  $\dot{\rho} = \mathcal L \rho$, where the Liouvillian superoperator $\mathcal L$ reads 
\begin{eqnarray}
	\mathcal L \rho = -i[H, \rho] + \sum_{\alpha}(J_{\alpha}\rho J_{\alpha}^{\dagger}-\frac{1}{2}\{J_{\alpha}^{\dagger}J_{\alpha}, \rho \}).
\end{eqnarray}
\begin{figure*}[htb] 
	\includegraphics[scale=0.28]{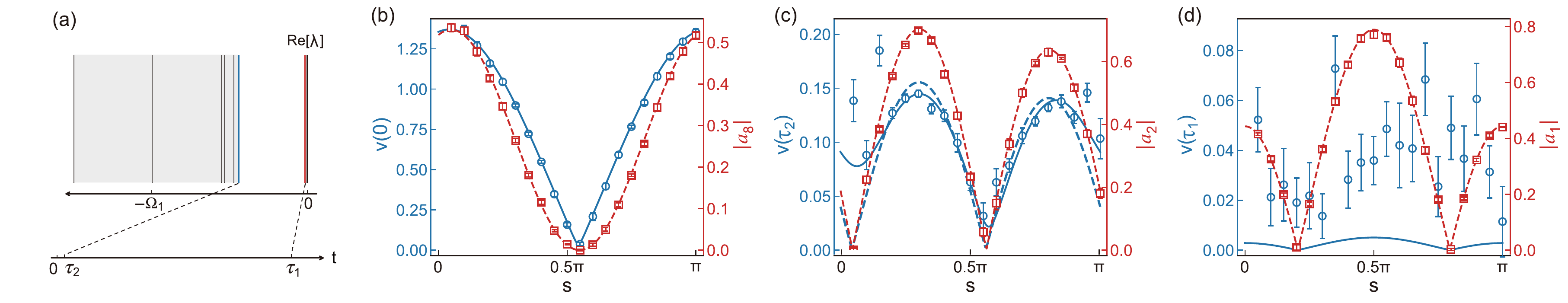}
	\caption{(a) Eigenspectra of $\mathcal L$ for three-level system and the corresponding time scale. The separation between $\lambda_1$ and $\lambda_2$ define a metastable manifold for times $\tau_2 \ll t \ll \tau_1$. (b)-(d) The relaxation speed $v$ (blue lines) of a rotated initial random state as a function of $s$ at time $t=0, \tau_2, \tau_1$, and the overlap $|a_8|, |a_2|, |a_1|$ (red lines), respectively. The blue dashed line in (c) corresponds to \(v(\tau_2) = \big\|\sum_{i=2}^{5} a_i \lambda_i e^{\lambda_i \tau_2} R_i\big\|\), which is a good approximation of the exact speed \(v(\tau_2) = \big\|\sum_{i=1}^{8} a_i \lambda_i e^{\lambda_i \tau_2} R_i\big\|\) (blue solid line).}
	\label{fig:overlap}
\end{figure*}
Here $H$ is the Hamiltonian of the system, and $J_{\alpha}$ are the quantum jump operators describing the dissipative effects of the open system. If Liouvillian operator is diagonalizable, its eigenmatrices and eigenvalues are defined via the relation that $\mathcal L R_i = \lambda_i R_i,\mathcal L^{\dagger} L_i = \lambda_i^* L_i$. $R_i(L_i)$ are the right (left) eigenmatrices of the Liouvillian superoperator $\mathcal L$, which form a basis for the space of matrices and can always be normalized in such a way $\mathrm{Tr}[L_i R_j ] =\delta_{ij}$. Since the dynamics generated by $\mathcal L$ are completely positive, the eigenvalues of $\mathcal L$ all have a nonpositive real part, $\mathrm{Re}[\lambda_i] \le 0$ \cite{Macieszczak2016a, Carollo2021, Minganti2019}. We sort the eigenvalues in such a way that $\lambda_0 > \mathrm{Re}[\lambda_1] \ge \mathrm{Re}[\lambda_2] \ge \mathrm{Re}[\lambda_3] \ge ...$, and $\lambda_0=0$. $\lambda_0$ or its eigenmatrix $R_0$ denotes the steady state $\rho_{ss}$, which is independent of any initial state $\rho_{in}$. The real parts of other eigenvalues $\lambda_{i\geq 1}$ indicate the relaxation rates of the eigenmodes $R_i$ and their corresponding life-time scales are $\tau_i \sim 1/|\text{Re}[\lambda_i]|$(see Fig. \ref{fig:overlap}(a)). The density matrix $\rho(t)$ hence can be expanded as the sum of all eigenmodes ($R_i$) of $\mathcal L$
\begin{equation}
	\rho_t = e^{\mathcal L t} \rho_{in}= \rho_{ss} +  \sum_{i=1}^{N} a_i e^{\lambda_i t} R_i,
\end{equation}
where coefficients $a_i = \mathrm{Tr}[L_i \rho_{in}]$ give the overlap of $L_i$ with $\rho_{in}$, and $N$ denotes the number of the decay modes. 

For long times \(t \gg \tau_2\), all decay modes with \(i\ge 2\) have essentially decayed, and the dynamics is dominated by the SDM. Consequently, the deviation from the steady state obeys \(\mathcal D(t) = \|\rho_t - \rho_{\mathrm{ss}}\| \propto |a_1 e^{\lambda_1 t}|\), where \(\|\cdot\|\) is a suitably chosen norm \cite{Kochsiek2022}. This shows that initial states with a smaller overlap \(|a_1|\) are indeed closer to the steady state in the long-time limit---a key mechanism used to explain the quantum ME. However, this criterion cannot capture transient relaxation information, such as why an initial state farther from equilibrium can relax faster than a closer one.

To get an intuitive interpretation of the quantum ME, we introduce the speed of the relaxation dynamics  \cite{Nava2025,Gaidash2025,Deffner2017a}:
\begin{eqnarray}
v(t) = \|\dot {\rho}_t \| = \lim_{\delta t \to 0} \frac{\| \rho_{t+\delta t} - \rho_{t} \|}{\delta t} = \|\sum_{i=1}^{N} a_i\lambda_i e^{\lambda_i t} R_i\|,
\end{eqnarray}
which is the scalar velocity of the velocity field $\dot{\rho}_t = \mathcal L \rho_t$ \cite{Nava2025}. Actually, the speed $v(t)$ given in Eq.~(3) is also
widely studied in quantum time limit bounds under various guises \cite{Pires2016,Deffner2013,Campo2013,Sun2015,Taddei2013}(see Supplemental Material I for details about distance and the derivation of bound). It is independent of the physical situation and considered quantities \cite{Deffner2017a,Deffner2017}, such as the distance $\mathcal D(t)$ studied here. The speed captures the combined influence of the initial state via the coefficients $a_i$ and the generator \(\mathcal{L}\) via the eigenvalues $\text{Re}[\lambda_i]$. Therefore, it does not require prior knowledge of the quantum state $\rho(t)$ and the steady state, making it more predictable and can quantify the disparity in relaxation speed among different initial states in transient regimes, which is exactly what we require to understand anomalous relaxation. 

Now we analyze the main factors influencing the relaxation speed on several important time scales and the results are presented in Fig. \ref{fig:overlap}(b-d) (See Supplemental Material II for details on relaxation speed and overlaps and Fig. S1). At time \(t=0\), if the relaxation rate of the fastest decay mode is significantly larger than the others, i.e., \(\lambda_N \gg \lambda_{i<N}\), then \(v(t=0) \propto |a_N \lambda_N|\), indicating that the initial relaxation speed is dominated by the fastest decaying mode (see Fig. \ref{fig:overlap}(b)). Meanwhile, at the intermediate times, the fast decay modes
can be neglected and considering \(|\lambda_{i\le2}| \gg |\lambda_1|\) so that the intermediate decay modes $R_{\{2,...,m\}}$ govern the relaxation speed, i.e.,  $v(t) \approx \|\sum_{i=2}^{m}a_i \lambda_i e^{\lambda_i t} R_i\|$. Especially, when \(t\gg 1/\tau_3\), as illustrated in Fig.~\ref{fig:overlap}(c), the relaxation speed is dominated by the second slowest decay mode, i.e., \(v(t) \propto |a_2 \lambda_2|\). Finally, when \(t\gg\tau_2\), where only the SDM remains, we get \(v(t) \propto |a_1 \lambda_1|\). Consequently, in the long-time limit, as illustrated in Fig. \ref{fig:overlap}(d), an initial state with a smaller overlap \(|a_1|\) lies closer to the stationary state but exhibits a slower relaxation speed. This indicates that the faster relaxation of the farther initial state is not caused by the smaller overlap \(|a_1|\) but rather by other overlaps. As a result, the trajectory crossing associated with the quantum ME does not occur in the long-time limit but at intermediate times (See Supplemental Material II for details on crossing time of quantum ME in Fig. S2). This highlights the importance of both the initial and the intermediate-time relaxation speeds for understanding the crossing behavior. 

\begin{figure*}[htb]
	\includegraphics[scale=0.3]{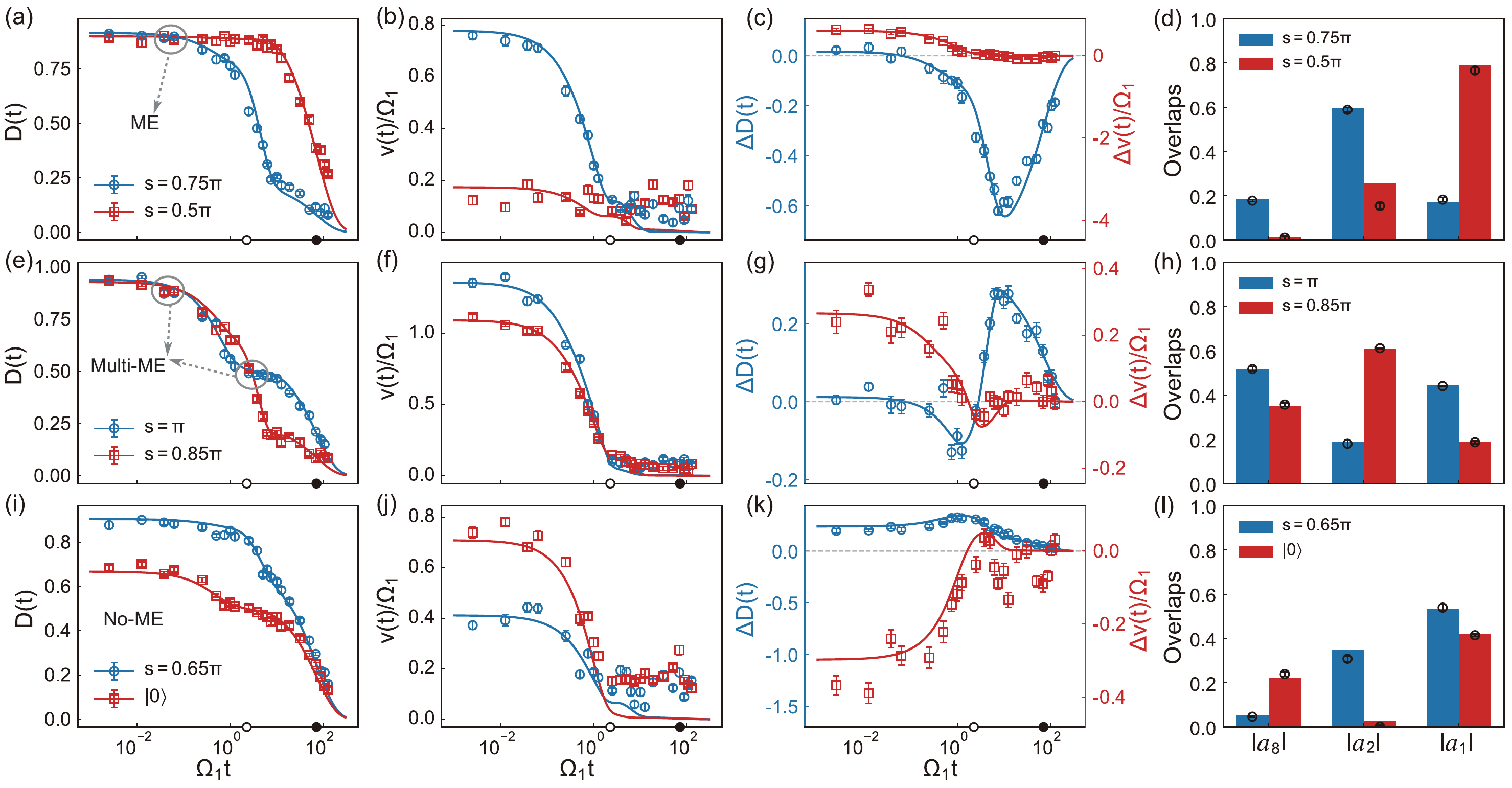}
	\caption{(a)-(d) Quantum ME. The distance $\mathcal D(t)$ (a) and relaxation speed $v(t)$ (b) as functions of time $t$ for different initial states $s = 0.75\pi$ (initial farther) and $s = 0.5\pi$ (initial closer). Open circle ($\circ$) in the horizontal axis is the time scale $\tau_2$ while the full circle ($\bullet$) is $\tau_1$. (c) The difference in distance $\Delta \mathcal D(t)=\mathcal D^f(t)-D^c(t)$ and speed $\Delta v(t) = v^f(t)-v^c(t)$ between these two states ('f' denotes the initial farther one and 'c' the closer one). (d) The overlaps $|a_1|$, $|a_2|$, $|a_8|$ of the initial states $s = 0.75\pi$ and $s = 0.5\pi$, respectively. (e)-(h) Quantum multi-ME for different initial states $s = \pi$ and $s =0.85 \pi$. (i)-(l) No quantum ME for different initial states $s = 0.65\pi$ and $|0\rangle$.}
	\label{fig:Distance}
\end{figure*}
\textit{Experimental results}—We consider the experimental demonstration in a qutrit system constructed by a trapped $^{40}\mathrm{Ca}^{+}$ ion (see Fig. \ref{fig:scheme}(e)). The single $^{40}$Ca$^+$ ion is confined in a blade-shaped linear Paul trap with trap frequency about $2\pi\times $ 1.2 MHz and  $2\pi\times $ 1.6 MHz in axial and radial direction respectively. The quantization axis of the system is defined by applying  a static magnetic field of 6 Gauss and lifts the degeneracy of the Zeeman sublevels (See Supplemental Material III for details on the experimental setup). The ground state $|0\rangle = |4^2S_{1/2},m_j=-1/2\rangle$ is coupled to two excited state $|1\rangle = |3^2D_{5/2},m_j=-5/2\rangle$ and $|2\rangle=|3^2D_{5/2},m_j=3/2\rangle$ by a bichromatic narrow linewidth 729 nm laser beam  with Rabi frequencies $\Omega_1$ and $\Omega_2$, respectively. The decay rates from $|1 (2)\rangle$ to $|0 \rangle$ is controlled by using a right-handed circularly polarized  854 nm laser beam, are denoted by $\gamma_{1(2)}$ and $\gamma_1\gg \gamma_2$ \cite{Zhang2025}. Then Hamiltonian of the electronic states of the ion is $H=\sum_{\alpha = 1,2}\Omega_{\alpha}/2(|0\rangle \langle \alpha|+|\alpha\rangle \langle 0|)$, and the jump operators are $J_{\alpha} = \gamma_{\alpha}|0\rangle \langle {\alpha}|$ ($\alpha = 1, 2$).

We initialize the ion in different initial pure states $\rho_{in}(s)=|s\rangle\langle s|$ by applying a unitary operation $U(s)$ \cite{Carollo2021,Zhang2025} on the ground state $| 0\rangle$ (See Supplemental Material IV for details on the construction of the initial state). Then we characterize the distance $\mathcal D(t)$ and the relation speed $v(t)$ using the Hilbert-Schimidt norm that \( \|\cdot\| = \sqrt{\operatorname{Tr}[(\cdot)^{\dagger}(\cdot)]} \). Other measures, such as trance distance \cite{Nava2025, Bao2025, Nava2024}, yield qualitatively similar outcomes (See Fig. S3 in Supplemental Material II). We plot $\mathcal D(t)$ and $v(t)$ over time $t$ for different initial state, respectively, under the same experimental parameters as Fig.\ref{fig:scheme}. Fig. \ref{fig:Distance} (a) reports a typical quantum ME that the curves of the distance cross just once between different two initial states ($s=0.75\pi,0.5\pi$) at a specific time $t<\tau_2$ and do not cross again for any times thereafter. Fig. \ref{fig:Distance} (b) gives the corresponding speeds, which shows that the further initial state ($s=0.75\pi$) relaxes faster than the closer one $s=0.5\pi$ till the system enters a metastable manifold (see Fig. \ref{fig:Distance}(c)). As shown in Fig.~\ref{fig:Distance}(d), the faster relaxation of \(s = 0.75\pi\) is due to its larger overlaps \(|a_8|\) and \(|a_2|\) compared to the state \(s = 0.5\pi\).

Fig.~\ref{fig:Distance}(e-h) presents results for another pair of initial states (\(s = \pi\) and \(s = 0.85\pi\)), which satisfy \(|a_1^f| > |a_1^c|\) ('f' denotes the farther initial states and 'c' the closer one). In contrast to the previous cases, they exhibit different dynamical behaviors. Although the state \(s = 0.85\pi\) is initially closer to the steady state than \(s = \pi\), the two quickly reverse order; however, a subsequent crossing occurs at a later time, restoring the initial ordering. Thus, in this case, the normal ME is avoided, and the initially closer state relaxes faster \cite{Chalas2024}. This multi-ME is driven by the overtaking of their relaxation speeds \(v(t)\) before the system relaxes into the metastable manifold spanned by the SDM $R_1$ and $\rho_{ss}$. As shown in Fig.~\ref{fig:Distance}(f,g), at the beginning, the speed of the initially farther state \(s = \pi\) is faster than that of the initially closer state \(s = 0.85\pi\), leading to the first overtaking. Later, however, it becomes slower than the state \(s = 0.85\pi\), resulting in a second crossing of the trajectories. The exchange of the speed order between the two states can be predicted by their overlaps \(|a_8|\) and \(|a_2|\): \(|a_8(s=\pi)| > |a_8(s=0.85\pi)|\) but \(|a_2(s=\pi)| < |a_2(s=0.85\pi)|\) (see Fig.~\ref{fig:Distance}(h)). It should be emphasized that the multi‑ME observed in our work is fundamentally different from the oscillatory distance behavior arising from complex eigenvalues of the SDM \cite{Chatterjee2024}, which has been experimentally observed in Ref.~\cite{Zhang2025}. In contrast, the multi‑ME we report occurs under a monotonically nonincreasing distance, where crossings are purely driven by the competition of different decay modes and not by any revival or oscillation of the distance. This distinctive feature makes our observed multi‑ME a more subtle and counterintuitive phenomenon that can't be captured by the complex‑eigenvalue scenario.

For the condition \(|a_1^f| > |a_1^c|\), there is another case as shown in Fig.~\ref{fig:Distance}(i-l) where no ME occurs: the initially closer state \(|0\rangle\) always relaxes faster than the farther one \(s = 0.65\pi\), and there is no crossing of their distance trajectories. This is because the overlap \(|a_8|\) of state \(|0\rangle\) is larger than that of state \(s = 0.65\pi\) (see Fig.~\ref{fig:overlap}(l)). Consequently, at the beginning, the speed of the initially farther state \(s = 0.65\pi\) is slower than that of the initially closer state \(|0\rangle\), causing the distance difference to increase (see Fig.~\ref{fig:overlap}(k)). Even though it later becomes faster than state \(|0\rangle\), it cannot overtake at any finite time, so no ME occurs.

\begin{figure}[htb]
	\includegraphics[scale=0.3]{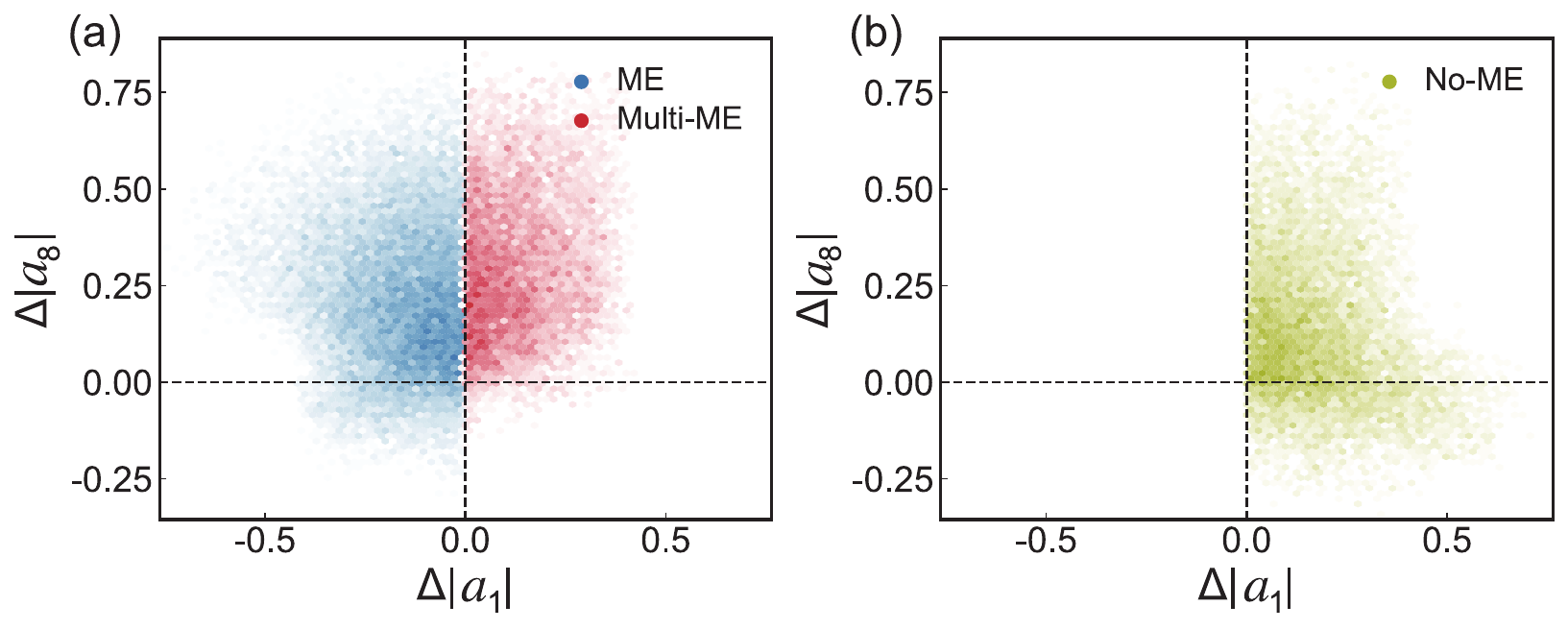}
	\caption{Phase diagram of the ME and Multi ME. The distribution of different types of ME as function of \(\Delta |a_8|\) and \(\Delta |a_1|\) ($\Delta |a_i| = |a_i^f|-|a_i^c|$) for (a) single crossing and two crossings and (b) no crossing.  Here we randomly generate 50000 pairs of arbitrary initial pure states. Here, the ME occurs with approximately 87.7\% probability when $|a_8^f| > |a_8^c|$ and $|a_1^f| < |a_1^c|$, while the multi-ME occurs with approximately 96.8\% probability when both $|a_8^f| > |a_8^c|$ and $|a_1^f| > |a_1^c|$ are satisfied.}
	\label{fig:distribition}
\end{figure}

We now  determine the phase diagram of the quantum ME. Based on the three cases above, we find that, on one hand, the overlaps of the fastest and the second-slowest decay modes play important roles in predicting the relaxation speed at the beginning and intermediate-time, respectively; on the other hand, the overlaps of the fastest and the SDM offer crucial help in determining the type of quantum ME (See Fig. S4 in Supplemental Material II). To verify these conclusions, we consider a more general scenario. We randomly generate 50000 pairs of arbitrary initial pure states (covering arbitrary $|s\rangle$ states) and statistically analyze the number of crossings in their distance trajectories as functions of \(\Delta |a_8|\) and \(\Delta |a_1|\) ($\Delta |a_i| = |a_i^f|-|a_i^c|$), with the results shown in Fig.~\ref{fig:distribition}. It can be seen that the quantum ME occurs when \(|a_1^f| < |a_1^c|\) (see Fig.~\ref{fig:distribition}(a)). When \(|a_1^f| > |a_1^c|\), typically, for most pairs of states the distance curves never cross, as shown in Fig.~\ref{fig:distribition}(b). However, as presented in Fig.~\ref{fig:distribition}(b), if in addition \(|a_8^f| > |a_8^c|\), multi-ME can occur.

\textit{Conclusions}—In summary, we have provided a transient picture of the quantum ME in Markovian open systems by introducing the concept of relaxation speed. Using a trapped‑ion platform, we experimentally observed multi‑ME in the regime where the farther initial state possesses a larger overlap with the slowest decay mode (SDM), a phenomenon that lies beyond the standard criterion. We further showed that the relaxation speed at initial and intermediate times is governed by the overlaps with the fastest and the intermediate decay modes, respectively. Together with the SDM overlap, these indicators allow us to find a phase diagram that shows both the both the occurrence and the type of the quantum ME. Moreover, the ability to predict which state relaxes fastest in different time intervals enables a relay strategy that maintains optimal relaxation speed at all times. This approach may benefit optimal engineering tasks, such as accelerating quantum battery charging or reducing computational costs in many‑body simulations where reaching the steady state is time‑consuming. Our work thus not only offers a transient picture of the quantum ME but also establishes a predictive framework for actively designing and optimizing relaxation processes in quantum technologies.


\textit{Acknowledgments}—Y.L.Z. acknowledges discussions with Yan-Yi Wang. This work was supported by the National Nature Science Foundation of China (Grant Nos. 12574553, 12574403) and the Innovation Research Foundation of NUDT. P.X.C. acknowledges the support by Quantum Science and Technology-National Science and Technology Major Project (Grant No. 2021ZD0301605). W.L. acknowledges support from the EPSRC through Grant No. EP/W015641/1. H.J. is supported by the NSFC (Grant No.11935006), the Science and Technology Innovation Program of Hunan Province (Grant No. 2020RC4047), National Key R\&D Program of China (No. 2024YFE0102400) and Hunan provincial major sci-tech program (2023ZJ1010).


%

\clearpage
\newpage
\onecolumngrid
\appendix

\setcounter{figure}{0}                     
\renewcommand{\thefigure}{S\arabic{figure}}

\renewcommand{\theequation}{\thesection S\arabic{equation}}
\setcounter{equation}{0}

\section{Supplemental Material for "Observation of quantum multi-Mpemba effect in a trapped-ion system"}

\subsection{I. Distance and the derivation of bound}
We define the Hilbert-Schmidt distance 
\begin{eqnarray}
	\mathcal{D}(\rho, \rho_{ss})=\|\rho_t-\rho_{ss}\|,
\end{eqnarray}
which is a geometric distance that signals the distinguishability between the steady state $\rho_{ss}$ and the time-dependent state $\rho_t$ starting from the initial state $\rho_{in}$.  Here $||A|| = \sqrt{\mathrm{Tr}[A^{\dagger}A}]$ is the Hilbert-Schimidt norm. Let us now calculate the changing rate of the distance, i.e., the geometric speed
\begin{eqnarray}
	v_{\mathcal D}(t) = |\dot{\mathcal D}|=|\frac{\mathrm{Tr}[(\rho-\rho_{ss})\dot{\rho}_t)]}{\|\rho-\rho_{ss}\|}|.
\end{eqnarray}
We can bound the such defined speed by using the Cauchy-Schwarz inequality for operators, $|\mathrm{Tr}[AB]|\le \|A\|\cdot \|B\|$. Then we obtain
\begin{eqnarray}
	v_{\mathcal D}(t)\le \frac{\|\rho_t-\rho_{ss}\|\| \dot{\rho}_t)\|}{||\rho_t-\rho_{ss}||}=\|\mathcal L\rho\|=v(t),
\end{eqnarray}
which means the $v(t)$ we defined in Eq. (3) in the letter is a kind of the quantum speed limit \cite{Pires2016,Deffner2013,Campo2013,Sun2015,Taddei2013}. However, we emphasize that there are multiple ways to characterize such a distance, and this non-uniqueness arises from the plethora of bona fide distance measures available in the convex space of quantum states \cite{Deffner2017a}. In all these cases, the distance is set by a Schatten-$p$-norm of the generator of the dynamics \cite{Deffner2013}: $v_p(t) \equiv \|\mathcal{L} \rho\|_p$, where $p$ is a positive real number. For $p=2$, this corresponds to the Hilbert-Schmidt distance; for $p=1$, the trace distance; and for $p = \infty$, the operator norm.


This relaxation speed of Eq. (3) undertakes the role of initial states and the generator operator $\mathcal L$. Thus it could quantify the disparity in relaxation speed among different initial states in transient regimes, which is exact we required to understand anomalous relaxation. At long-time limit that $t>>\tau_2=1/|\mathrm{Re}[\lambda_2]|$, the influence of the short-life decay modes on the relaxation process becomes very small, while the SDM is still relevant and keeps the system away form the steady state till $t>>\tau_1=1/|\mathrm{Re}[\lambda_1]|$. This means, that for long times, one has $\rho(t) -\rho_{ss} \approx c_1 e^{\lambda_1 t}$ and the changing rate of the distance can be rewrite as 
\begin{eqnarray}
	v_{\mathcal D}(t)= v(t)\approx |c_1\lambda_1|||R_1||e^{\lambda_1 t}.
\end{eqnarray} 
It indicates the system approaches steady state with the speed upper bound $v(t)$, proceeding along the direction of SDM $R_1$. 

\subsection{II. The quantum ME, relaxation speed and the overlaps}
To state more exactly the existence of a ME, we need to know the transient dynamics in more detail in the view of relaxation speed. First, we need to establish the connection between the relaxation speed and the overlaps. Here we randomly generate 50000 pairs of arbitrary initial pure states. Among them, 2,000 are the arbitrary $|s\rangle$ states discussed in the main text. We statistically analyze their relaxation speed at different time as the function of the overlaps. As shown in Fig. S1(a), at time \(t = 0\), the relaxation speed is almost liner with the overlap of the fastest decay mode \(|a_8|\), meaning that the initial state with a larger overlap on the fastest decay mode will relax faster at the beginning. While on the timescale \(\tau_1\), \(v(\tau_1) \propto |a_1|\) (see Fig. S1(a)). Considering there are several eigenvalues $\lambda_{3,4,5}$ near with the eigenvalue $\lambda_2$ (see Fig. \ref{fig:overlap}(a)), at the time scale of $\tau_2$, the speed actually is determined by the the decay modes $R{2\sim5}$. However, for the arbitrary $|s\rangle$ we generated in the main text, on the timescale \(\tau_2\), as illustrated in Fig. S1(b), we deed have \(v(t=2\tau_2) \propto |a_2|\). 
\begin{figure}[htb]
	\includegraphics[scale=0.45]{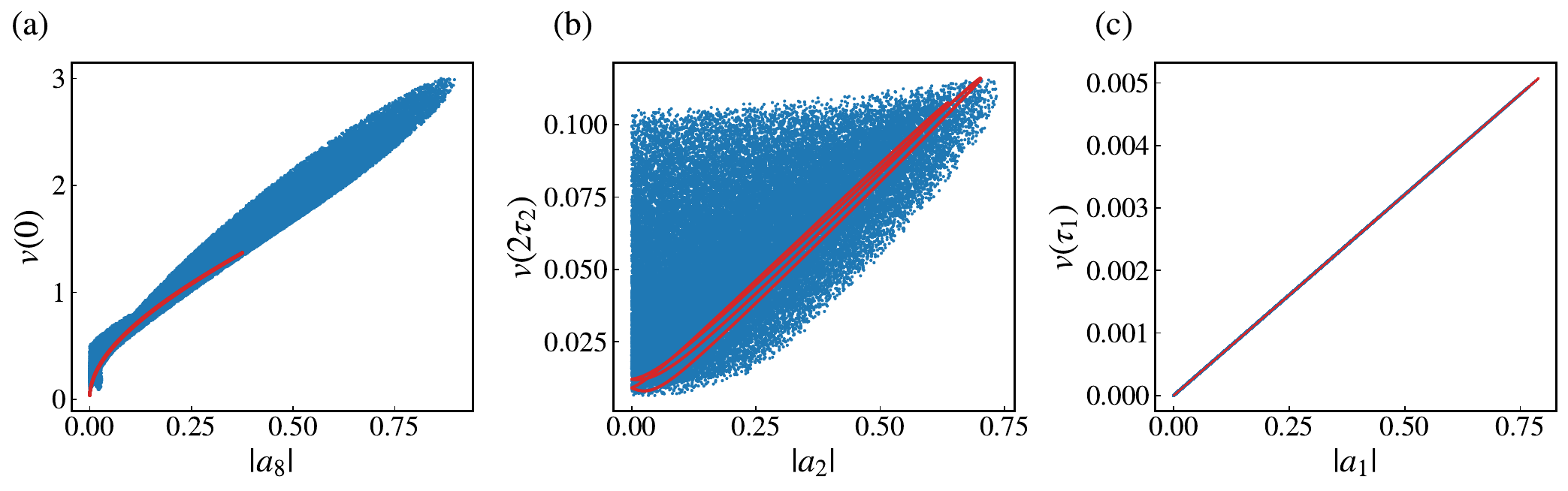}
	\caption{The connection between the relaxation speed at times \(t = 0\), \(2\tau_2\), \(\tau_1\) and the overlaps \(|a_1|\), \(|a_2|\), \(|a_8|\) for random initial pure states. Here we randomly generate 50000 pairs of arbitrary initial pure states. Among them, 2,000 are the arbitrary $|s\rangle$ states discussed in the main text.}
\end{figure}

Normally, to determine whether the quantum Mpemba effect (ME) occurs, one often needs to scan all possible initial state and calculate or measure their evolution trajectories, and compare their distances as a function of time to observe crossing. In Fig. S2, we statistically analyzed the distribution of crossing times for arbitrary pairs of initial states, and find that there are two crossings when $|a_1^f| > |a_1^c|$, as illustrated in Fig. S2 (b). The crossing times for both the normal quantum ME and the multi-ME indeed occur before the long-time limit., i.e., $t\ll \tau_1$. In Fig. S3, we also choose the trance distance as the measurement and yield qualitatively similar outcomes as the Hilbert-Schimidt distance shown in Fig. \ref{fig:Distance}. 
\begin{figure}[htb]
	\includegraphics[scale=0.45]{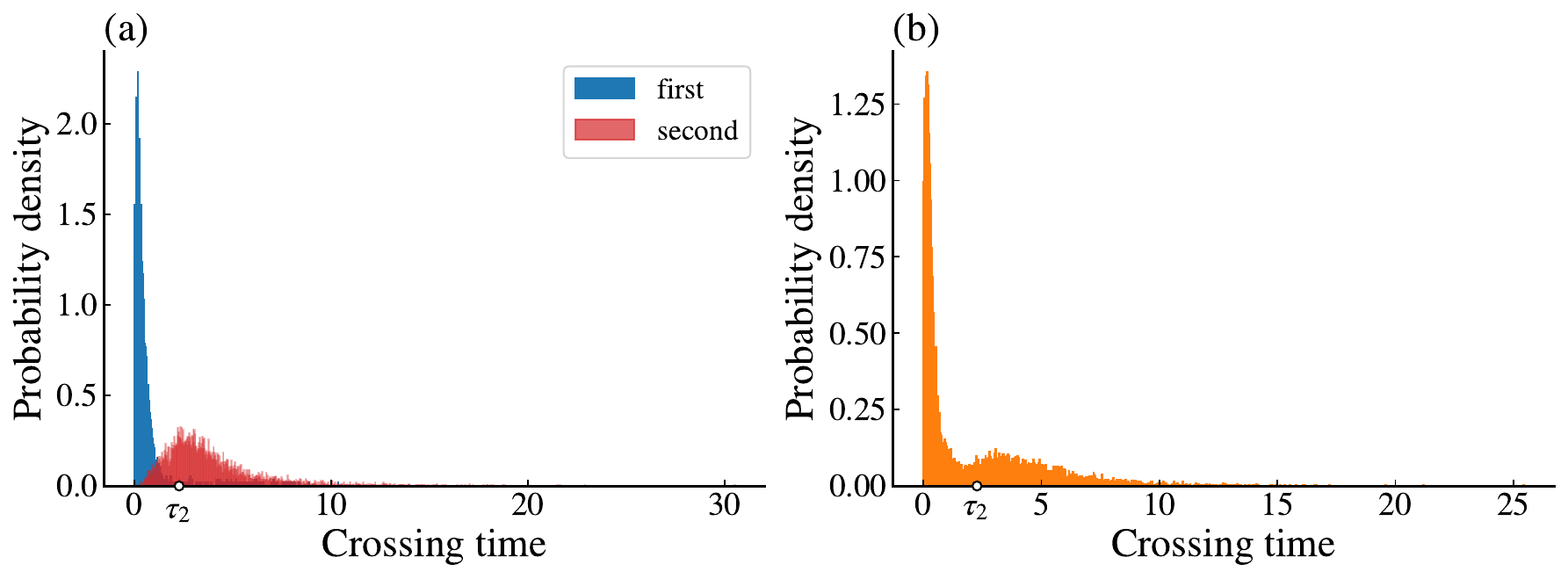}
	\caption{The distribution of crossing times for arbitrary pairs of initial states for the Multi-ME (a) and ME (b) for random initial pure states. Blue: the first crossing time, red: the second crossing time. Here we randomly generate 50000 pairs of arbitrary initial pure states.}
\end{figure}

\begin{figure}[htb]
	\includegraphics[scale=0.3]{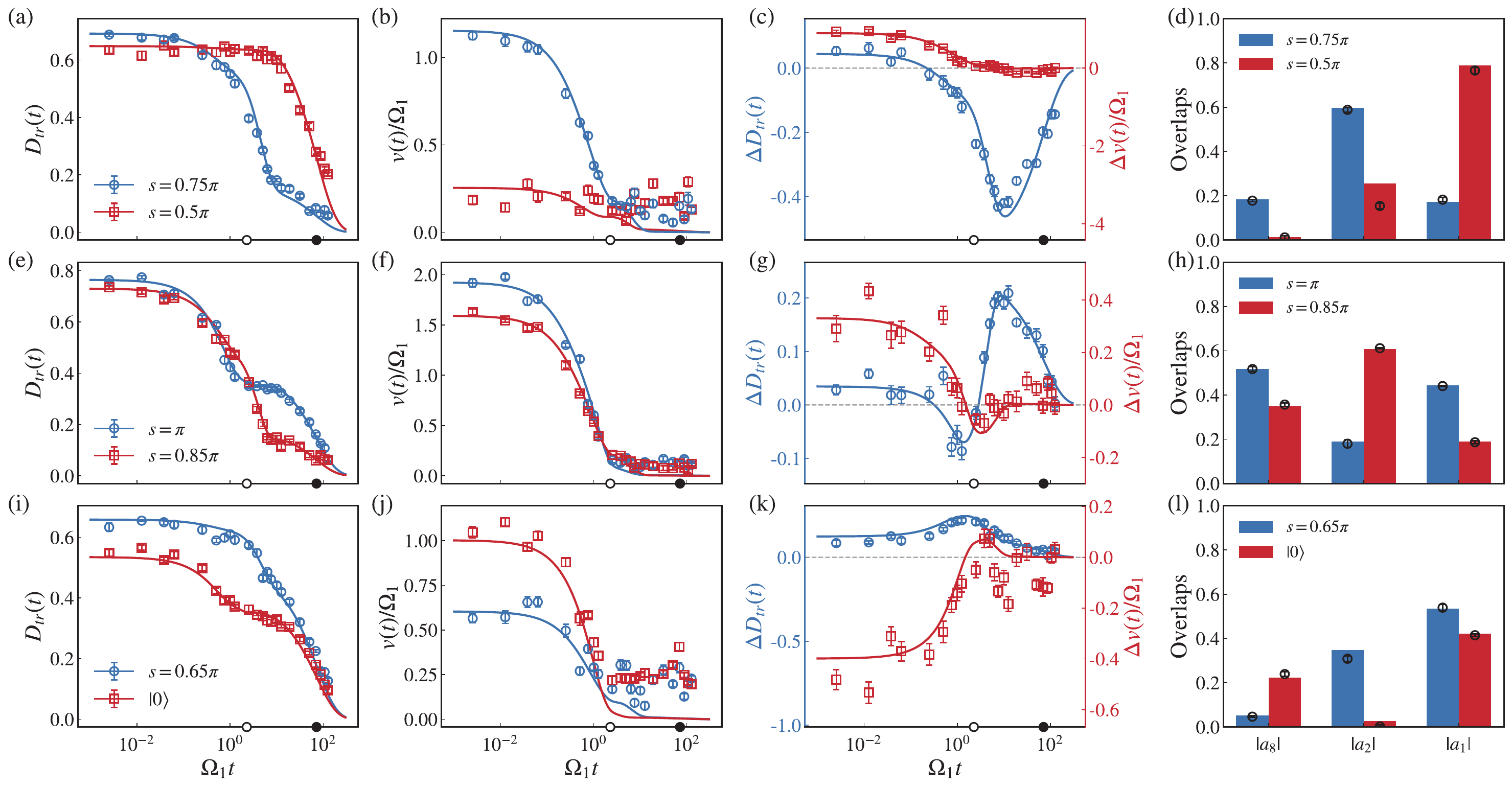}
	\caption{(a)-(d) Quantum ME. The trace distance $\mathcal D_{tr}(t)$ (a) and the corresponding relaxation speed $v_{tr}(t)$ (b) as functions of time $t$ for different initial states $s = 0.75\pi$ (initial farther) and $s = 0.5\pi$ (initial closer). Open circle ($\circ$) in the horizontal axis is the time scale $\tau_2$ while the full circle ($\bullet$) is $\tau_1$. (c) The difference in distance $\Delta \mathcal D_{tr}(t)=\mathcal D^f(t)-D^c(t)$ and speed $v_{tr}(t) = v^f(t)-v^c(t)$ between the states $s =0.75 \pi$ and $s =0.5\pi$. (d) The overlaps $|a_1|$, $|a_2|$, $|a_8|$ of the initial states $s = 0.75\pi$ and $s = 0.5\pi$, respectively. (e)-(h) Quantum multi-ME. The distance $\mathcal D_{tr}(t)$ (e) and relaxation speed $v_{tr}(t)$ (f) as functions of time $t$ for different initial states $s = \pi$ and $s =0.85 \pi$. (g) The difference in distance $\Delta \mathcal D_{tr}(t)$ and speed $\Delta v_{tr}(t)$ between these two states. (h) The overlaps $|a_1|$, $|a_2|$, $|a_8|$ of the initial states $s = \pi$ and $s = 0.85\pi$, respectively. (i)-(l) No quantum ME. The distance $\mathcal D_{tr}(t)$ (i) and relaxation speed $v_{tr}(t)$ (j) as functions of time $t$ for different initial states $s = 0.65\pi$ and $|0\rangle$. (k) The difference in distance $\Delta \mathcal D_{tr}(t)$ and speed $\Delta v(t)$ between the states $s = 0.65\pi$ and $|0\rangle$. (l) The overlaps $|a_1|$, $|a_2|$, $|a_8|$ of the initial states $s = 0.65\pi$ and $|0\rangle$, respectively.}
\end{figure}

\begin{figure}[htb]
	\includegraphics[scale=0.45]{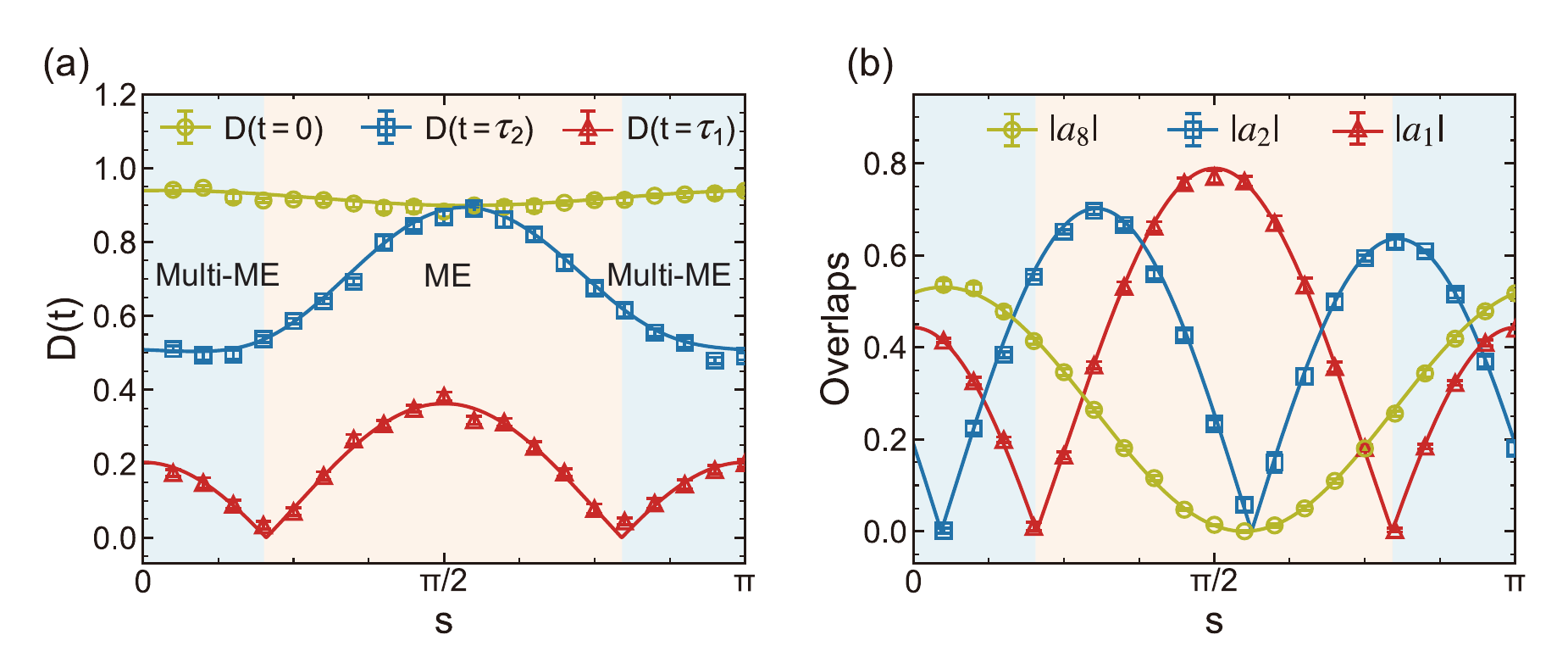}
	\caption{(a) The distance $\mathcal D(t)$ of a rotated initial random state as a function of $s$ at time $t=0$, $\tau_2$ and $\tau_1$, respectively. (b) Overlaps $|a_1|$, $|a_2|$ and $|a_8|$ of a rotated initial random state as a function of $s$.}
\end{figure}
Although such a procedure is possible, we usually use a much simpler way to test for a ME---the overlap $a_1$. It is the only term retaining information on the initial condition in the long-time limit, implying that it can determine the existence of the ME for a given pair of initial state ("f" denotes the farther one and "c" the closer one) if $|a_1^f| < |a_1^c|$, then at long times, $\mathcal D^f(t) < \mathcal D^c(t)$. Indeed it can imply that there exists some finite time $t > 0$ after which the system that started from further is closer to the target steady state and thus provides a sufficient condition to determine the existence of the ME. Unfortunately, this criterion can not predict multi-ME which could happen when a farther initial state instead exhibits a larger SDM overlap, i.e., $|a_1^f| > |a_1^c|$. In this case the overlaps with the fastest decay mode and the second slowest decay mode play important roles. For arbitrary initial states $|s\rangle$, the phase diagram of the quantum ME is shown in Fig. S4(a). As illustrated in Fig. S4(a, b), in the ME regime, the farther initial state has a smaller overlap $|a_1|$ but a larger $|a_8|$, which leads to a crossing of the distance. Thus the distance order change once. In the multi-ME regime, the farther initial state has larger overlaps $|a_{1,8}|$ but a smaller $|a_2|$.  Notably, $|a_8|$ determines the initial relaxation speed, $|a_2|$ governs the intermediate speed (or the occurrence of the second crossing), and $|a_1|$ dictates the final ordering toward the steady state. As a result, the distance order change twice and multi-ME happens (see Fig. S4(a)). This means that, together with the overlaps $|a_1|$, $|a_2|$, and $|a_8|$, one can explain the occurrence of the quantum ME and multi-ME.

\subsection{III. The experimental setup}

Here we take a single $^{40}$Ca$^+$ ion  to construct the Markovian open quantum system.  A blade-shaped linear Paul trap with an axial  and radial trap frequency of $2\pi \times 1.2 $  MHz and $2\pi \times 1.6$ MHz is used to confine a single $^{40}$Ca$^+$ ion. The quantization axis is defined by applying  a static magnetic field of 6 Gauss with an angle of 45 degrees with respect to the trap axis. The qutrit system is encoded in the sublevels $|S_{1/2}, m_j = -1/2 \rangle$, $|D_{5/2}, m_j = -5/2 \rangle$ and $|D_{5/2}, m_j = 3/2 \rangle$, The S and D states are coupled by a narrow linewidth bichromatic  729 nm laser beam, which is generated by injecting two frequency components into an acousto-optic modulator via an arbitrary waveform generator. The decay rate of the D states is adjusted by tuning the power and setting polarization of the 854 nm laser beam, here the right-circularly polarized 854 nm beam is chosen to make sure the decay rate of state  $|D_{5/2},m_j=3/2\rangle$ is 3 orders of magnitude lower than that of state $|D_{5/2},m_j=-5/2\rangle$.  Before applying the coupling 729 nm laser beam, the ion is prepared to motional ground state by using EIT cooling with a mean phonon number about 0.05. At the end of the experiment, a 397 nm laser beam  is applied to generate the fluorescence in S state and a photomultiplier tubes (PMT) is used to collect the photons for determining the probability of S state, and the probabilities of the two D states are obtained by the same method but with a swapping operation on S and D states before applying the 397 nm detection beam.

\subsection{VI. Construction of the initial state and qutrit State Tomography}

We detail the construction of the initial state $\rho_{\mathrm{s}}$ by taking advantage of the unitary transformation $U$, which can be written as 	
\begin{equation}
	U = \exp\!\big[-i s (|\phi_1\rangle\langle \phi_2| + |\phi_2\rangle\langle \phi_1|)\big] |\phi_1\rangle\langle 0|,
\end{equation}
where  $s = \arctan\!\big(\sqrt{|\alpha_1/\alpha_2|}\,\big)$ is the angle calculated by the eigenvalues $\alpha_{1(2)}$ of the  left eigenmatrix $L_1$.  In the experiment, the qutrit (three-level) unitary $U$ can  be realized by using a series of the  elementary two-level rotations \cite{Nielsen2011}, but all of these operations should have a proper phase so that the global phase fulfill the requirement of the initial state. Moreover,  any phase shift  relative to the spectator level has impact on the final results \cite{Ringbauer2022}.

Fortunately, this obstacle can be overcome by a more efficient decomposition that involves the virtual phase gates in the  two-level rotations. Consider an arbitrary target qutrit  state $|\psi\rangle = U|0\rangle = (a, b, c)^\mathrm{T}$ with $|a|^2+|b|^2+|c|^2=1$. The experimental unitary can be decomposed as $U = Mm$ \cite{Nielsen2011}, where
\begin{align}
	M &= 
	\begin{pmatrix}
		\frac{a}{\sqrt{|a|^2+|b|^2}} & \frac{b^*}{\sqrt{|a|^2+|b|^2}} & 0 \\[4pt]
		\frac{b}{\sqrt{|a|^2+|b|^2}} & -\frac{a^*}{\sqrt{|a|^2+|b|^2}} & 0 \\[4pt]
		0 & 0 & 1
	\end{pmatrix}, \\
	m &= 
	\begin{pmatrix}
		\sqrt{|a|^2+|b|^2} & 0 & c^* \\[4pt]
		0 & 1 & 0 \\[4pt]
		c & 0 & -\big(\sqrt{|a|^2+|b|^2}\big)^*
	\end{pmatrix}.
\end{align}
Note that matrices $A $ and $ B $ can be express by using phase and rotation gates with 
\begin{align}
	M &= P_{01}(\alpha) Z_{01}(\beta) R_{01}(\gamma,\pi/2) Z_{01}(\delta) \nonumber \\
	&= P_{01}(\alpha) R_{01}^z(\beta+\delta) R_{01}(\gamma,\pi/2-2\delta), \\
	m &= P_{02}(\alpha') Z_{02}(\beta') R_{02}(\gamma',\pi/2) Z_{02}(\delta') \nonumber \\
	&= P_{02}(\alpha') Z_{02}(\beta'+\delta') R_{02}(\gamma',\pi/2-2\delta'),
\end{align}
and the gate is defined as 
\begin{align}
	P_{01}(x) &= 
	\begin{pmatrix}
		e^{ix} & 0 & 0 \\
		0 & e^{ix} & 0 \\
		0 & 0 & 1
	\end{pmatrix}, &
	P_{02}(x) &= 
	\begin{pmatrix}
		e^{ix} & 0 & 0 \\
		0 & 1 & 0 \\
		0 & 0 & e^{ix}
	\end{pmatrix}, \\
	Z_{01}(x) &= 
	\begin{pmatrix}
		e^{-ix} & 0 & 0 \\
		0 & e^{ix} & 0 \\
		0 & 0 & 1
	\end{pmatrix}, &
	Z_{02}(x) &= 
	\begin{pmatrix}
		e^{-ix} & 0 & 0 \\
		0 & 1 & 0 \\
		0 & 0 & e^{ix}
	\end{pmatrix}, \\
	R_{01}(\theta,\phi) &= 
	\begin{pmatrix}
		\cos\frac{\theta}{2} & -i e^{-i\phi}\sin\frac{\theta}{2} & 0 \\[4pt]
		-i e^{i\phi}\sin\frac{\theta}{2} & \cos\frac{\theta}{2} & 0 \\[4pt]
		0 & 0 & 1
	\end{pmatrix}, \\
	R_{02}(\theta,\phi) &= 
	\begin{pmatrix}
		\cos\frac{\theta}{2} & 0 & -i e^{-i\phi}\sin\frac{\theta}{2} \\[4pt]
		0 & 1 & 0 \\[4pt]
		-i e^{i\phi}\sin\frac{\theta}{2} & 0 & \cos\frac{\theta}{2}
	\end{pmatrix}.
\end{align}
The parameters determined from the matrix elements are
\begin{align}
	\gamma &= 2\cos^{-1}(M_{11}), &
	\alpha &= \tfrac{1}{2}\big[\arg(M_{11}) + \arg(M_{22})\big], \\
	\beta &= \tfrac{1}{2}\big[\arg(M_{21}) - \arg(M_{11})\big], &
	\delta &= -\alpha + \arg(M_{22}) - \beta, \\
	\gamma' &= 2\cos^{-1}(m_{11}), &
	\alpha' &= \tfrac{1}{2}\big[\arg(m_{11}) + \arg(m_{33})\big], \\
	\beta' &= \tfrac{1}{2}\big[\arg(m_{31}) - \arg(m_{11})\big], &
	\delta' &= -\alpha' + \arg(m_{33}) - \beta'.
\end{align}
Here, the phase gates $P$ and $Z$ are shifted backward to implement the required phase shifts between qutrit levels \cite{Ringbauer2022,McKay2017}. The full unitary then becomes
\begin{align}
	U &= P_{01}(\alpha) P_{02}(\alpha') Z_{01}(\beta+\delta) Z_{02}(\beta'+\delta') \nonumber \\
	&\quad \times R_{01}(\gamma,\phi) R_{02}(\gamma',\phi'),
\end{align}
with $\phi = \alpha' - (\beta'+\delta') + \frac{\pi}{2} - 2\delta$ and $\phi' = \frac{\pi}{2} - 2\delta'$.

We further simplify the implementation by noting that the phase gates can be commuted through the subsequent dynamics and absorbed into redefinitions of the driving phases, since detection zero sensitive to global phases \cite{McKay2017}. Consequently, only the two rotations $R_{01}$ and $R_{02}$ need to be physically implemented, significantly simplifying the operation and improving the fidelity of the initial state.

After preparing the initial state, the dynamical evolution of the system and also the state tomography pulses are applied.  Note that the accumulated  phase shifts in the state preparation process  needs to be accounted in the dynamical evolution and tomography process, but we can use the same strategy to deal with the phase, and  the corresponding adjustments to the dynamical and tomography operations can be simply expressed as
\begin{equation}
	\mathcal{L}(\Omega_1, \Omega_2) \rightarrow \mathcal{L}(\Omega_1 e^{i\phi}, \Omega_2 e^{i\phi'}).
\end{equation}\
The state tomography for a qutrit system requires  9 bases, here we choose the bases as: $\{|0\rangle, |1\rangle, |2\rangle, (|0\rangle\pm|1\rangle)/\sqrt{2}, (|0\rangle\pm i|1\rangle)/\sqrt{2}, (|0\rangle\pm|2\rangle)/\sqrt{2}, (|0\rangle\pm i|2\rangle)/\sqrt{2}, (|1\rangle\pm|2\rangle)/\sqrt{2}, (|1\rangle\pm i|2\rangle)/\sqrt{2}\}$. Since  only the $S$ (ground) and $D$ (metastable) manifolds can be distinguished in the detection process, all the  measurement basis needs to be rotated to $|0\rangle$ before the detection. The tomography rotation $\mathcal{T}(\vartheta_1 e^{i\phi_{L1}}, \vartheta_2 e^{i\phi_{L2}})$ is realized via $\pi$ or $\pi/2$ pulses on the transitions $|0\rangle\leftrightarrow|1\rangle$ and $|0\rangle\leftrightarrow|2\rangle$, with accumulated phases $\phi_{L1} = \alpha' - 2(\beta+\delta) - (\beta'+\delta')$ and $\phi_{L2} = \alpha - 2(\beta'+\delta') - (\beta+\delta)$, respectively. Finally, the density matrix at each time stamp  is reconstructed from the projective measurements using maximum-likelihood estimation, and the error bars are obtained by Monte Carlo simulations \cite{PhysRevA.66.012303,Qin2023}.

\end{document}